# A Miniature Cold-Atom Frequency Standard

[1]V.Shah, [2]M. Mescher, [2]R. Stoner, [3]V. Vuletic, [1]R. Lutwak

[1]Symmetricom Technology Realization Centre, Beverly, MA, 01915

[2]The Charles Stark Draper Laboratory, Cambridge, MA, 02139

[3]MIT – Research Laboratory of Electronics, Cambridge, MA, 02139

**Atomic sensors employing cold-atom technology enable unprecedented accuracy and resolution for next-generation atomic clocks, magnetometers, gravimeters, and gyroscopes. To date, however, the size and complexity of cold atom systems have prevented their deployment in practical applications outside of large research laboratories. Here we demonstrate a low power, palm-top, and fully integrated cold atom system that functions as an atomic clock with a stability of 2 parts in $10^{11}$ at 1s. This work demonstrates the feasibility of developing compact, robust, and portable devices based on laser cooled atoms.**

In the two decades since the first experimental demonstration of laser cooling of alkali atoms, many promising applications have been demonstrated[1–4]. Today, the world's most accurate atomic clocks are based on ultra-cold atoms because they provide a long interrogation time and because the magnitude of the perturbations and systematic offsets that limit conventional atomic clocks are reduced by several orders of magnitude[5]. While these laboratory demonstrations have generated considerable excitement, the technology has yet to be deployed outside of controlled environments, due to their cost, complexity, and environmental sensitivity. Miniature trapping geometries have been previously demonstrated, but these are typically fiber-coupled to large and complex laser systems. In this work we have demonstrated a miniature cold-atom physics package, occupying less than 75 cm$^3$, *including all necessary laser systems*. The device is demonstrated as a miniature cold-atom frequency standard (MCAFS), with frequency stability (Allan Deviation) of $\sigma_y(\tau) < 2 \times 10^{-11} \tau^{-1/2}$ /√Hz, though other applications in magnetometry, gravimetry, and rotation sensing are readily accessible from the core technology. The relative simplicity and small size of the MCAFS has been achieved with the development and integration of novel components and techniques for laser cooling and trapping, including a compact magneto-optic trap[6] (MOT) operated by a single laser, and a chip-scale vapor cell.

Figure 1 shows a photograph (i) and cutaway view (ii) of the MCAFS. From left to right, the MCAFS consists of a (a) a temperature stabilized diode laser, (b) an optical shutter, (c) beam expander, (d) photodetector, and (e) MOT cell. The MOT cell contains a conical reflector (f) and is contained within a pair of magnetic-field coils (g).

All necessary laser wavelengths are generated by a single laser device, which is a 15 mW 852 nm high-efficiency distributed Bragg reflector (DBR) laser, optimized for narrow spectral linewidth (< 2 MHz) and high-modulation efficiency (> 3 GHz). The injection current of the laser is modulated to generate a second order sideband at 8.93 GHz to substitute for a separate "repumping" laser.

A sample of the laser light is directed towards a miniature dichroic atomic vapor absorption laser lock[7] (DAVLL) assembly (v), for stabilization of the laser wavelength and calibrated hopping of the laser frequency to different points during the clock cycle. In this fashion, the single laser sequentially plays the role of laser cooling, optical pumping, and resonant detection. The DAVLL unit is a modified chip-scale atomic clock (CSAC) physics package based on polarimetry using a single photodiode (instead of the conventional balanced detection using two photodiodes). Long term accuracy of the DAVLL is maintained by periodically calibrating to the resonance fluorescence signal observed from the cold atoms during the cooling stage of the clock cycle.

The main part of the laser beam passes though an optical shutter, which is comprised of a stack of four liquid crystal display (LCD) polarization rotators, each sandwiched between a pair of linear polarizers with combined optical switching speed of <100 microseconds, light extinction ratio of 55 dB, and forward loss of less than 3 dB. Following the shutter, a pair of beam expansion lenses and a quarter wave retarder produce a 1.6 x 1.8 cm$^2$ circularly polarized light beam.

The technique of using conical/pyramidal mirror for laser cooling, developed by Lee. et al.[8], provides significant simplification of the optical setup and relaxes the otherwise tight constraints on laser alignment. The three pairs of opposing circularly-polarized beams necessary for MOT operation are all derived from a single large-diameter input beam by a conical mirror. The 1.6 cm diameter conical mirror in MCAFS was fabricated by electroforming of copper onto a highly polished, diamond turned Ni-P plated stainless steel mandrel.

The MOT cell itself is made from nonmagnetic stainless steel with a helium impermeable glass window in the front. In addition to the conical mirror, the cell contains an electrically heated cesium metal dispenser (AMD) and a non-evaporable getter (NEG) pump. For these early experiments, the background pressure inside the UHV chamber (<1x10$^{-9}$ Torr) is maintained by a small 2 l/s ion pump connected to the main chamber by a 1/8$^{th}$ inch diameter stainless tube, though efforts are underway to eliminate the pump and maintain vacuum solely with the NEG. Typically the AMD is activated for five to ten minutes every few days to replenish the cesium vapor background pressure. To avoid clock perturbation due to the magnetic fields generated by the electrical activation, the AMD is operated synchronous with the laser cooling sequence of the clock cycle.

The MCAFS illustrated in Figure 1 does not include a optical isolator to prevent back reflections from the MOT cone from perturbing the laser frequency. We have found that the level of perturbations due to optical feedback does not impact the efficiency of laser cooling, which is sufficient for many cold atom applications. For clock operation, however, where the signal-to-noise of the fluorescence signal is critically dependent on frequency stability, we have found it to be necessary to add a Faraday optical isolator to the system (not shown in Figure 1). Figure 2 shows clock data taken with another MCAFS prototype, nearly identical in construction to the one shown in Fig. 1, except including a relatively large cylindrical Faraday optical isolator (1.5 (d) x 1.5 (h) cm). Unfortunately, millimeter-scale optical isolators with low insertion loss are unavailable at the cesium wavelength of 852 nm. An MCAFS based on $^{87}$Rb atoms is currently under development, for the primary reason that miniature optical isolators are available at the wavelength of the principal transition in rubidium (780 nm).

The Figures 2 (a) and (b) show the Cesium microwave resonance at 9.192 GHz, interrogated with the Ramsey and Rabi techniques, respectively. When operated as a closed-loop clock, typical stability is as shown in Figure 2(c). While the theoretical (shot-noise) limit of MCAFS has the potential to achieve stability below $1\times10^{-12}/\sqrt{\tau}$, we have found experimentally that the clock stability is currently limited by laser frequency noise.

**Methods:** When operating as a clock, the basic principles of the MCAFS are similar to that of existing clocks based on laser cooled atoms that rely on microwave Ramsey spectroscopy, such as large-scale cold-atom atomic "fountain" clocks, except that the cold atoms are "dripped" from the trap[9,10] for interrogation rather than "tossed" upwards[11].

The clock operates cyclically as follows:

A beam of laser light tuned slightly to the red of F=4 → F'=5 D2 transition in $^{133}$Cs atoms shines on to a highly polished copper cone. The cone reflects the light in all directions to cool and trap roughly ten million cesium atoms at a mean temperature of 100 μK in a magneto-optic trap (MOT). Once the atoms are trapped (in about 100 ms) the magnetic fields are turned off and the laser is adiabatically detuned to produce additional cooling. The atoms are then optically pumped to the lower F=3 ground state hyperfine level by turning off the microwave modulation of the laser. The laser light is then switched off and the atoms are released from the trap. In the dark, as the atom cloud is freely expanding and falling under gravity, microwave pulses, nearly resonant with the |F=3, m$_f$=0⟩ to |F=4, m$_f$=0⟩ ground state hyperfine frequency of $^{133}$Cs, are applied using a small microwave loop. Typically, the microwave power is adjusted to produce two "π/2" pulses, each 1 ms long, separated by 15 ms. The laser is then tuned to the peak of the F=4 → F'=5 D2 transition and the optical switch is reopened. From the initial fluorescence signal from the cold atoms, the number of atoms transferred to the F=4 ground state is determined. The number of atoms that are transferred to the F=4 state has a well known functional dependence on the microwave frequency, the so-called "Ramsey Resonance," shown in Figure 2 (a). A clock is created by locking the frequency of the microwaves to the central peak, which exhibits stability as shown in Figure 2(c).

**Acknowledgements** The authors would like to thank Armand Martinis for his work on fabricating the compact MOT cell. The authors would also like to sincerely thank Brian Timmons, Nicole Byrne, John LeBlanc, Jonathan Bernstein, Mitch Hansberry, Fran Rogomentich, Don Emmons and Mike Garvey. This project was funded the Defense Advanced Research Projects Agency (DARPA), Contract # N66001-09-C-2057.

Distribution Statement "A" (Approved for Public Release, Distribution Unlimited)

\*\* **Disclaimer: "The views expressed are those of the author and do not reflect the official policy or position of the Department of Defense or the U.S. Government."**

1. Kasevich, M.A., Riis, E., Chu, S. & DeVoe, R.G. rf spectroscopy in an atomic fountain. *Phys. Rev. Lett.* **63**, 612 (1989).

2. Vengalattore, M. *et al.* High-Resolution Magnetometry with a Spinor Bose-Einstein Condensate. *Phys. Rev. Lett.* **98**, 200801 (2007).

3. Peters, A., Chung, K.Y. & Chu, S. Measurement of gravitational acceleration by dropping atoms. *Nature* **400**, 849-852 (1999).

4. Gustavson, T.L., Bouyer, P. & Kasevich, M.A. Precision Rotation Measurements with an Atom Interferometer Gyroscope. *Phys. Rev. Lett.* **78**, 2046 (1997).

5. Hall, J.L., Zhu, M. & Buch, P. Prospects for using laser-prepared atomic fountains for optical frequency standards applications. *J. Opt. Soc. Am. B* **6**, 2194-2205 (1989).

6. Monroe, C., Swann, W., Robinson, H. & Wieman, C. Very cold trapped atoms in a vapor cell. *Phys. Rev. Lett.* **65**, 1571 (1990).

7. Corwin, K.L., Lu, Z.-T., Hand, C.F., Epstein, R.J. & Wieman, C.E. Frequency-Stabilized Diode Laser with the Zeeman Shift in an Atomic Vapor. *Appl. Opt.* **37**, 3295-3298 (1998).

8. Lee, K.I., Kim, J.A., Noh, H.R. & Jhe, W. Single-beam atom trap in a pyramidal and conical hollow mirror. *Opt. Lett.* **21**, 1177-1179 (1996).

9. Monroe, C., Robinson, H. & Wieman, C. Observation of the cesium clock transition using laser-cooled atoms in a vapor cell. *Opt. Lett.* **16**, 50-52 (1991).

10. Magalhães, D.V., Müller, S.T., Bebeachibuli, A., Alves, R.F. & Bagnato, V.S. Comparative short-term stability for a Cs beam and an expanding cold atomic cloud of Cs used as atomic frequency standards. *Laser Physics* **16**, 1268-1271 (2006).

11. Wynands, R. & Weyers, S. Atomic fountain clocks. *Metrologia* **42**, S64-S79 (2005).

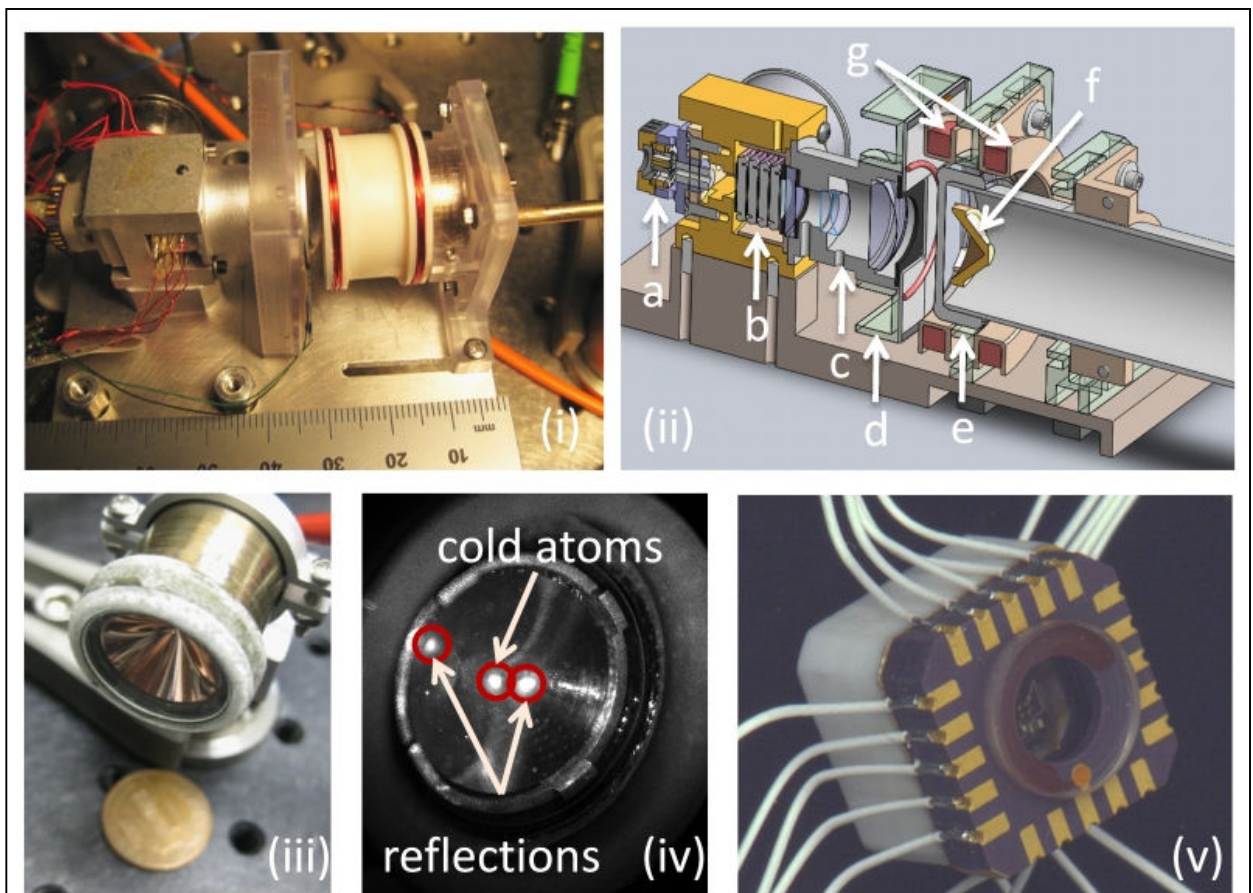

**Figure 1:** (i) a picture of MCAFS, a miniature cold atom clock based on ultra cold atoms. (ii) a 3D cross-sectional view of the MCAFS. (a) packaged diode laser, (b) optical shutter, (c) expansions lenses and quarter wave plate, (d) photodiodes for fluorescence collection, (e) UHV chamber, (f) copper cone, (g) magnetic field coils for MOT. (iii) A picture of the miniature UHV chamber used in MCAFS, seen with the installed copper cone. (iv) A picture of the UHV chamber with roughly 30 million ultra cold atoms suspended in a MOT. (v) A CSAC type miniature DAVLL assembly used in MCAFS to stabilize the laser frequency.

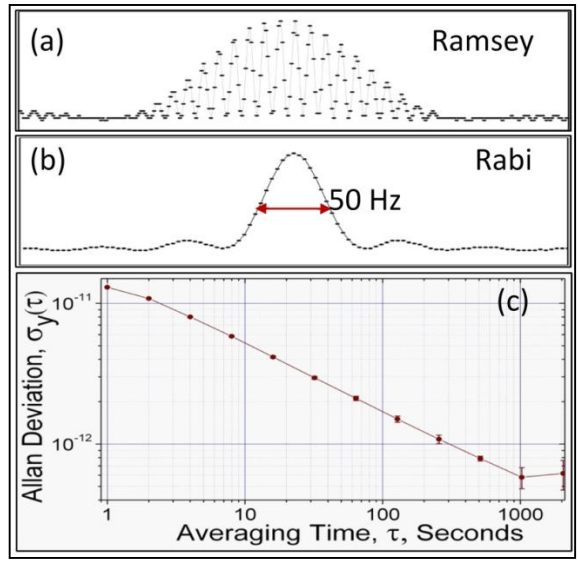

**Figure 2:** Microwave resonance seen by sweeping the LO frequency. (a) Two pulse microwave Ramsey interrogation. Frequency sweep = 3 kHz. (b) Rabi interrogation with 1 one 16 ms microwave pulse. Frequency sweep = 500 Hz. (c) Clock stability in terms of Allan deviation. The stability was measured without enclosing the MCAFS inside magnetic shields; we believe as a result the clock drifted over times scales longer than 1000 seconds.